## This is the preliminary version of the JASIST submission

# Applying centrality measures to impact analysis: A coauthorship network analysis

**Erjia Yan**[1]**, Ying Ding**

*School of Library and Information Science, Indiana University, 1320 East 10th Street, Bloomington, IN 47405-3907. E-mail: {eyan, dingying}@indiana.edu*
*Tel: (812) 606-8091*

## Abstract

Many studies on coauthorship networks focus on network topology and network statistical mechanics. This article takes a different approach by studying micro-level network properties, with the aim to apply centrality measures to impact analysis. Using coauthorship data from 16 journals in the field of library and information science (LIS) with a time span of twenty years (1988-2007), we construct an evolving coauthorship network and calculate four centrality measures (closeness, betweenness, degree and PageRank) for authors in this network. We find out that the four centrality measures are significantly correlated with citation counts. We also discuss the usability of centrality measures in author ranking, and suggest that centrality measures can be useful indicators for impact analysis.

## 1. Introduction

Social network analysis has developed as a specialty in parallel with scientometrics since the 1970s, examples as Hubbell's measure of sociometric status, Bonacich and Freeman's measure of centrality, Coleman's measure of power, and Burt's measure of prestige (Friedkin, 1991). The last decade has witnessed a new movement in the study of social networks, with the main focus moving from the analysis of small networks to those with thousands or millions vertices, and with a renewed attention to the topology and dynamics of networks (Newman, 2001a). This new approach has been driven largely by the improved computing technologies which allow us to gather and analyze data in large scales, which makes it possible to uncover the generic properties of social networks (Albert & Barabási, 2002).

Coauthorship network, an important form of social network, has been intensively studied in this movement (Newman, 2001a; Newman, 2001b; Barabási, Jeong, Neda, Ravasz, Schubert, & Vicsek, 2002; Nascimento, Sander & Pound, 2003; Kretschmer, 2004; Liu, Bollen, Nelson, & Sompel, 2005; Yin, Kretschmer, Hanneman, & Liu, 2006; Vidgen, Henneberg, & Naude, 2007;

[1] Corresponding author

Rodriguez & Pepe, 2008). Most of these researches focus on macro-level network properties, which informs us about the "likely performance of the social structure that arises out of the physics of its connections; the actors embedded in the network may well be completely unaware of this structure" (Yin et al., 2006, p. 1600), such as mean distance, clustering coefficient, component and degree distribution; yet not enough attention is paid to micro-level structure, which informs us about "the differential constraints and opportunities facing individual actors that shape their social behavior" (p. 1600), such as the power, stratification, ranking, and inequality in social structures (Wasserman & Faust, 1994). This article shows an example of studying micro-level structure by applying centrality measures to coauthorship network. Using twenty years (1988-2007) data from 16 journals in the field of library and information science, we construct an evolving coauthorship network, with the focus of testing the usability of centrality measures in impact analysis.

## 2. Backgrounds

Centrality analysis is not new to sociology. In a ground laying piece, Freeman (1977) developed a set of measures of centrality based on betweenness. In a follow-up article, Freeman (1979) elaborated four concepts of centrality in a social network, which have since been further developed into degree centrality, closeness centrality, betweenness centrality, and eigenvector centrality. Some influential research on this topic includes: the relationship between centrality and power (Hackman, 1985; Bonacich, 1987; Ibarra, 1993; Ibarra & Andrews, 1993), relationship between salience and psychological centrality (Stryker & Serpe, 1994), centrality on choices and behaviors (Verplanken & Holland, 2002), centrality within family (Crosbieburnett, 1984), organization networks (Boje & Whetten, 1981; Paullay, Alliger, & Stoneromero, 1994), groups and classes (Everett & Borgatti, 1999), as well as classroom social positions (Farmer & Rodkin, 1996).

Centrality has also been applied to journal impact analysis. Using journal data from Institute for Scientific Information (ISI), Bollen, Rodriguez, and Van De Sompel demonstrated how a weighted version of the popular PageRank algorithm can be used to obtain a metric that reflects prestige. They contrasted the rankings of journals according to ISI impact factor and weighted PageRank, and discovered that they both significant overlaps and differences. Leydesdorff (2007) applied betweenness centrality to 7,379 journals included in the Journal Citation Reports, and found that betweenness centrality is shown to be an indicator of the interdisciplinarity of journals. Dellavalle, Schilling, Rodriguez, Van de Sompel, and Bollen (2007) studied dermatology journals using weighted PageRank algorithm which assigned greater weight to citations originating in more frequently cited journals. They found that the weighted PageRank algorithm provided a more refined measure of journal status and changes relative dermatology journal rankings.

As for coauthorship networks, several articles have also applied centrality measures to coauthorship network analysis. Mutschke (2003) employed centrality to the coauthorship network of digital libraries research. Liu et al. (2005) applied centrality analysis to coauthorship of Joint Conference on Digital Libraries (JCDL) research community, and compared three kinds of centrality measures with the ranking of JCDL program committee membership, and discovered that betweenness centrality performed best among the three centrality measures. Estrada and Rodriguez-Velazquez (2005) proposed a new centrality measure that characterizes the participation of each node in all subgraphs in a network. They found that this centrality displayed useful and desirable properties, such as clear ranking of nodes and scale-free characteristics. Chen (2006) used betweenness centrality to highlight potential pivotal points of paradigm shift of scientific literature over time. Yin et al. (2006) applied three centrality measures to COLLNET community coauthorship network. Vidgen, Henneberg, and Naude (2007) applied five centrality measures (degree, betweenness, closeness, eigenvector, flow betweenness, and structural holes) to rank information system community. Similarly, Liu et al. (2007) applied betweenness centrality to the weighted coauthorship network of nature science research in China. These articles applied centrality measures to bibliometric analysis; some stepped further in ranking the authors through different centrality measures and compared them with bibliometric measures (Liu et al., 2005; Yin et al., 2006). But they did not elaborate the relation of centrality with citation for author ranking, or the usability of centrality in author's impact evaluation. In this article, we try to fill this gap by constructing an evolving coauthorship network and verifying the usability of centrality measures in scientific evaluation, and discussing its strengths and limitations.

## 3. Methodology

### 3.1. Centrality Measures

In this study, we apply three classic centrality measures (degree centrality, closeness centrality and betweenness centrality) and PageRank to the coauthorship network.

Degree centrality. Degree centrality equals to the number of ties that a vertex has with other vertices. The equation of it is as following where $d(n_i)$ is the degree of $n_i$ :

$$C_D(n_i) = d(n_i)$$

Generally, vertices with higher degree or more connections are more central to the structure and tend to have a greater capacity to influence others. For some authors with high degree, it is because they co-authored with many authors in a single paper, rather than co-authored in many papers.

PageRank. PageRank is initially proposed by Page and Brin (1998), who developed a method for assigning a universal rank to web pages based on a weight-propagation algorithm called PageRank. A page has high rank if the sum of the ranks of its backlinks is high. This idea is captured in the PageRank formula as follows:

$$PR(p) = (1-d)\frac{1}{N} + d\sum_{i=1}^{k}\frac{PR(p_i)}{C(p_i)}$$

where N is the total number of pages on the Web, d is a damping factor, C(p) is the outdegree of p, and $p_i$ denotes the inlinks of p. Thus, PageRank is actually the directed weighted degree centrality.

Closeness centrality. A more sophisticated centrality measure is closeness (Freeman, 1979) which emphasizes the distance of a vertex to all others in the network by focusing on the geodesic distance from each vertex to all others. Closeness can be regarded as a measure of how long it will take information to spread from a given vertex to others in the network (Yin et al., 2006). Closeness centrality focuses on the extensivity of influence over the entire network. In the following equation, $\mathrm{C_c}(n_i)$ is the closeness centrality, and $d(n_i, n_j)$ is the distance between two vertices in the network.

$$\mathrm{C_c}(n_i) = \sum_{i=1}^{N} \frac{1}{d(n_i, n_j)}$$

Betweenness centrality. Betweenness centrality is based on the number of shortest paths passing through a vertex. Vertices with a high betweenness play the role of connecting different groups. In the following formula, $g_{jik}$ is all geodesics linking node j and node k which pass through node i; $g_{jk}$ is the geodesic distance between the vertices of j and k.

$$C_B(n_i) = \sum_{j,k \neq i} \frac{g_{jik}}{g_{jk}}$$

In social networks, vertices with high betweenness are the brokers and connectors who bring others together (Yin et al., 2006). Being between means that a vertex has the ability to control the flow of knowledge between most others. Individuals with high betweenness are the pivots in the network knowledge flowing. The vertices with highest betweenness also result in the largest increase in typical distance between others when they are removed.

### 3.2. Data processing

We choose the top 16 leading LIS journals based on ratings by deans and directors of North American programs accredited by the ALA (Nisonger & Davis, 2005) as well as on Journal Citation Reports (JCR) data for the years 1988-2007.We excluded from the rankings non-LIS journals such as *MIS Quarterly*, *Journal of the American Medical Informatics Association*, *Information Systems Research*, *Information & Management*, and *Journal of Management Information Systems*. Meanwhile, since during this time period, some journals have changed their names, we also include these sources into our data set. These 16 journals are: *Annual Review of Information Science and Technology*, *Information Processing and Management*, *Scientometrics*, *Journal of the American Society for Information Science and Technology* (*Journal of the American Society for Information Science*), *Journal of Documentation*, *Journal of Information Science*, *Information Research*, *Library and Information Science Research*, *College and Research Libraries*, *Information Society*, *Online Information Review* (*Online and CD-ROM Review*, *On-Line Review*), *Library Resources and Technical Services*, *Library Quarterly*, *Journal of Academic Librarianship*, *Library Trends*, *Reference and User Services Quarterly*.

We download the twenty-year data of these 16 journals from the database of Web of Science. There are 22,380 documents in all, in which we just focus on articles and review articles, and the number for them is 10,344 (54 anonymous articles are excluded).

## 4. Results and analysis

### 4.1. An overview

After downloading the data from the ISI Web of Science, we extract the coauthorship network through Network Workbench (NWB, 2006). Since some authors used middle name initials for some of their papers, while not for the other papers. We combine the same authors manually by their affiliation information (e.g. we combine Meho, L and Meho, LI into one author in the network), and export the network to Pajek in gaining the largest component, mean distance, largest distance and clustering coefficient, showing in TABLE 1.

TABLE 1. Summary statistics for LIS coauthorship network

| | Values |
|---|---|
| Number of papers | 10,344 |
| Number of authors | 10,579 |
| Papers per author | 2.40 |
| Authors per paper | 1.80 |
| Average collaborators | 2.24 |
| Largest component | 20.77% |
| mean distance | 9.68 |
| Clustering coefficient | 0.58 |

There are 10,579 authors in this network, in which average author writes 2.40 papers, average paper has 1.80 authors, and average author collaborate with 2.24 authors. These are relatively low values comparing to the coauthorship networks of biology and physics constructed by Newman (2001b), who found that papers per author, authors per paper, and average collaborators for biology coauthorship network are 6.4, 3.75 and 18.1, and for physics coauthorship network the values are 5.1, 2.53 and 9.7. This is due to two factors: first, library and information scientists are less collaborative than biologists and physicists. In our data set, only 39 authors have collaborated with more than 18 authors, which is the median number of collaborators for biology coauthorship network. Second, biologists and physicists tend to collaborate more frequently and more widely due to their research requirements. It is not unusual for papers published on biological journals to have more than 10 authors, but this is quite rare for LIS articles. TABLE 2 shows the accumulative distribution of papers and authors.

TABLE 2. Accumulative distribution of papers and authors

| Year | Papers | Authors | Year | Papers | Authors |
|---|---|---|---|---|---|
| 1988 | 392 | 545 | 1998 | 4724 | 4832 |
| 1989 | 797 | 1012 | 1999 | 5271 | 5338 |
| 1990 | 1201 | 1462 | 2000 | 5802 | 5884 |
| 1991 | 1638 | 1890 | 2001 | 6322 | 6378 |
| 1992 | 2039 | 2262 | 2002 | 6891 | 6941 |
| 1993 | 2428 | 2671 | 2003 | 7456 | 7461 |
| 1994 | 2835 | 3066 | 2004 | 8073 | 8106 |
| 1995 | 3281 | 3486 | 2005 | 8773 | 8843 |
| 1996 | 3750 | 3913 | 2006 | 9535 | 9713 |
| 1997 | 4234 | 4357 | 2007 | 10344 | 10579 |

The number of papers and authors increases gradually. The two curves fit $y = 363.95t^{1.08}$ and $y = 492.00t^{0.98}$ $(t = 1, 2, 3, \cdots)$ respectively, with $R^2 = 0.9973$ and $R^2 = 0.9932$. This result indicates that the number of papers and authors will increase approximately with these curves in the coming years. Their evolving graphs are showing in FIG. 1.

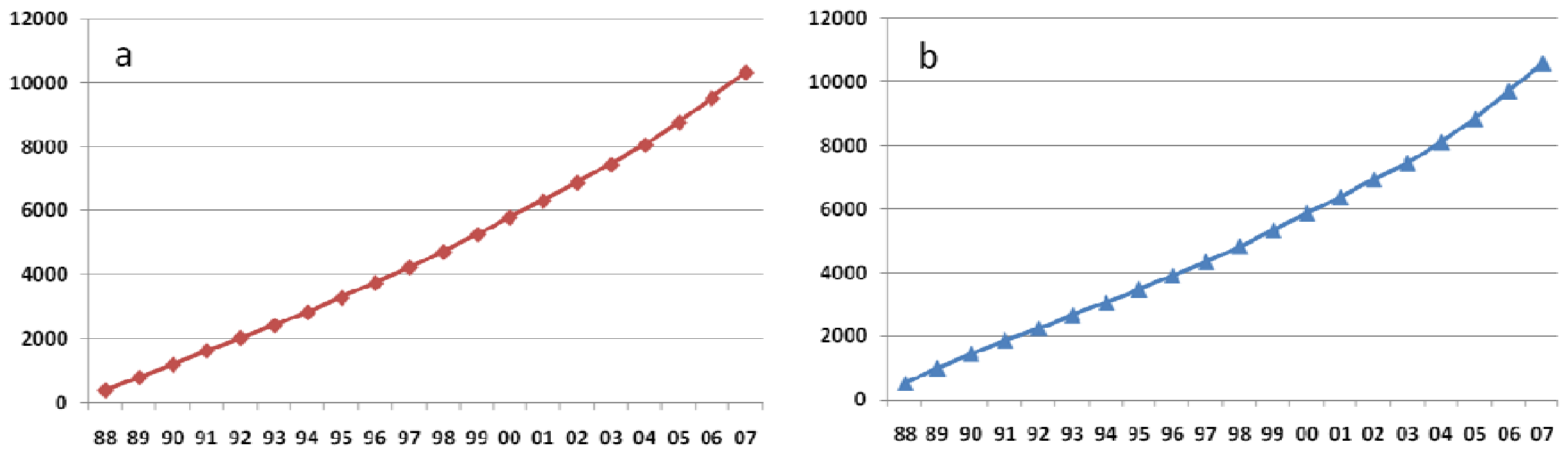

FIG. 1. Yearly accumulative distribution of papers (a) and authors (b).

Similar to observations from previous research on coauthorship networks, the LIS coauthorship network is not a single connected graph. The largest component of the network has 2,197 authors, taking about 20% of the total authors in the network. Nascimento, Sander, and Pound (2003) reported that the largest component in SIGMOD's coauthorship graph has about 60% of all authors. In the four coauthorship networks studied by Newman (Newman, 2001b), Medline has the largest component, with 92.6% of all the authors, while NCSTRL has the smallest largest component, containing 57.2% of all authors. After some comparison studies on coauthorship networks, Kretschmer (2004) suggests that the largest components usually have a ratio of more than 40% of all the authors. Our research only includes 16 journals which potentially cut some collaboration ties between authors; meanwhile, the nature of disciplines under study also affect this ratio: more authors would be involved if it is an experimental research, thus disciplines like biology and physics would have a bigger size of largest component.

TABLE 3. Properties of the evolving LIS coauthorship network from 1988 to 2007

| Year | Number of authors | Number of papers | Mean collaborators | Largest component | | |
|---|---|---|---|---|---|---|
| | | | | Size | Ratio% | Avg. distance |
| 1988-1992 | 2,262 | 2,039 | 1.70 | 46 | 2.26 | 2.49 |
| 1988-1997 | 4,357 | 4,234 | 1.76 | 91 | 2.15 | 5.30 |
| 1988-2002 | 6,941 | 6,891 | 1.91 | 646 | 9.37 | 9.54 |
| 1988-2007 | 10,579 | 10,344 | 2.24 | 2197 | 21.24 | 9.68 |

TABLE 3 shows the properties of the evolving LIS coauthorship network. Each author averagely has more collaborators, from 1.70 collaborators in the 1988-1992 period to 2.24 in 1988-2007 period. The increased mean collaborator means that authors collaborate more widely in recent years, which indicates that this field is becoming more collaborative.

The values of the largest component exhibit some diverse facts. In their study on mathematics and neuro-science coauthorship networks, Barabási et al. (2002) found that the mean distance of the mathematics coauthorship network decreased from 16 in 1991 to 9 in 1998, and the mean distance of the neuro-science coauthorship network decreased from 10 in 1991 to 6 in 1998. However, the mean distance of the LIS coauthorship increases from 2.49 in 1992 to 9.68 in 2007. The discrepancy is due to the fact that more new authors are involved in this field each year, but their collaboration pattern is simple and collaboration scope is limited comparing to neuro-science. Although LIS is increasingly becoming more collaborative, yet it has not arrived at its "phase transition" (Barabási, 2003) where authors collaborate with each other much more frequently and more widely, and from the perspective of network analysis, the mean distance will decrease after that phase.

### 4.2. Applying centrality measures to author ranking

Historically, most research on coauthorship network analysis focuses on overall topology of networks, whereas few researches has been done to discover individual properties, fewer on the relationship between citations and centrality measures. In this study, we calculate four centrality measures for authors in the largest component through Pajek. Their frequency distributions are shown in FIG. 2.

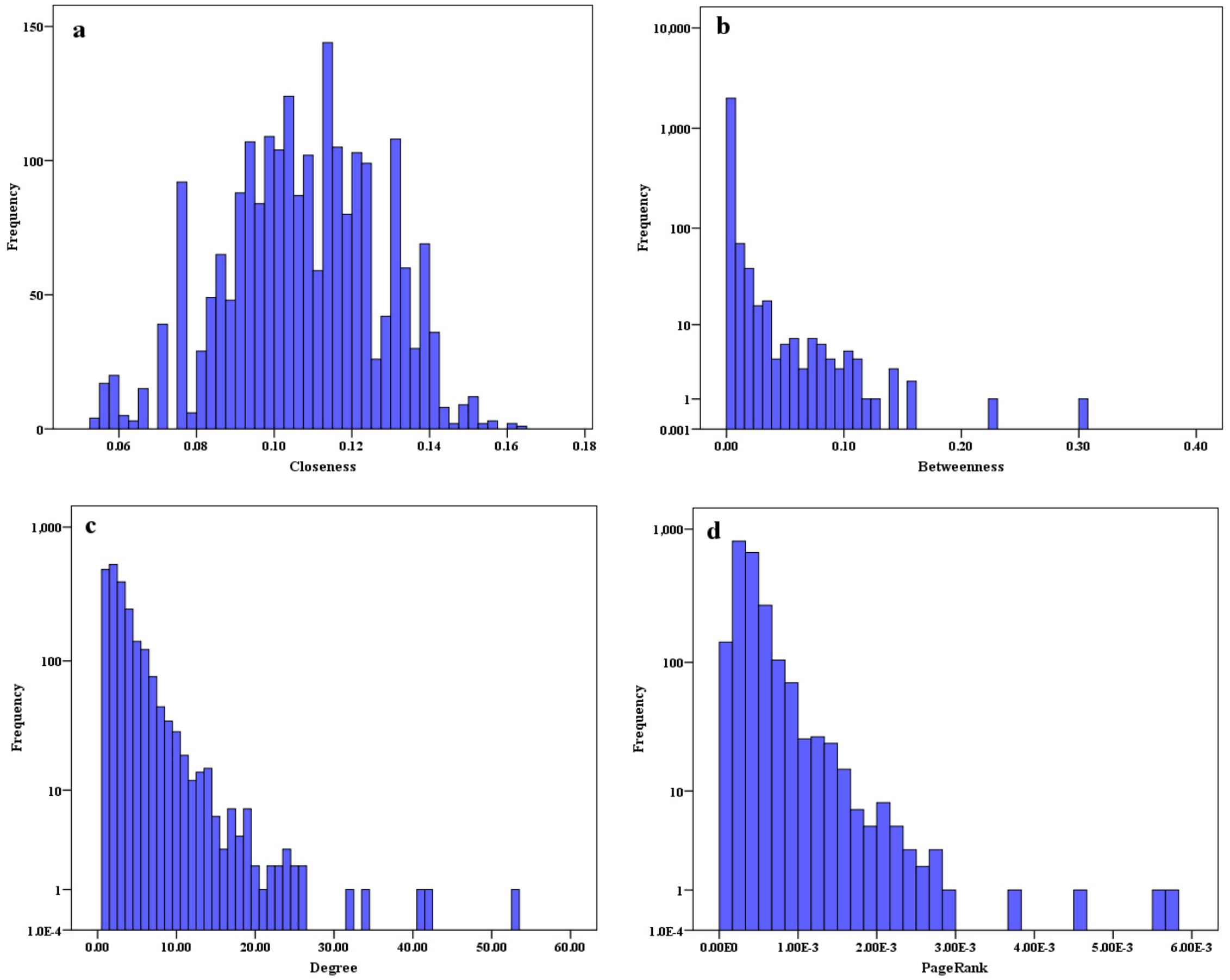

FIG. 2. Frequency distribution of closeness (a), betweenness (b), degree (c) and PageRank (d) centrality.

The frequency of betweenness centrality, degree centrality, and PageRank follows power-law distribution where most authors have low centrality values while a few authors have high centrality values. On the other hand, the distribution of closeness centrality follows the normal curve. The power-law distribution of degree centrality also indicates that this coauthorship network has scale-free character (Barabási & Albert, 1999): the relationship between degree and its frequency probability matches the curve: $p(k) = 1.1788k^{-2.1514}$, with $R^2$=0.9186. This result is also consistent with Price's network of citations (Price, 1965). He quoted a value of α = 2.5 to 3 for the exponent of his network. Other relevant researches on scale-free network also confirmed Price's assumption (Newman, 2003).

TABLE 4 through TABLE 7 show the top 30 authors based on closeness centrality, betweenness centrality, degree centrality, and PageRank calculated with the coauthorship network of 1988-1992, 1988-1997, 1988-2002, and 1988-2007 respectively. Authors appeared consecutively in the four time slices are marked in bold font, and authors appeared in three time slices are marked with italic font.

TABLE 4. Top 30 authors based on closeness centrality

| Rank | 1988-1992 | 1988-1997 | 1988-2002 | 1988-2007 | Rank | 1988-1992 | 1988-1997 | 1988-2002 | 1988-2007 |
|---|---|---|---|---|---|---|---|---|---|
| 1 | **Willett, P** | **Willett, P** | *Spink, A* | *Spink, A* | 16 | Rohde, NF | Rao, IKR | Schamber, L | Kantor, P |
| 2 | **Wood, FE** | *Bawden, D* | *Ellis, D* | **Willett, P** | 17 | Miquel, JF | Walker, S | Ozmutlu, HC | Kretschmer, H |
| 3 | *Bawden, D* | **Wood, FE** | Ford, N | *Ellis, D* | 18 | Straub, D | *Saracevic, T* | Greisdorf, H | *Bawden, D* |
| 4 | Cringean, JK | Beaulieu, M | *Losee, RM* | Ford, N | 19 | Beath, CM | Robertson, S | Robins, D | Jansen, BJ |
| 5 | Manson, GA | *Ellis, D* | **Willett, P** | Wilson, TD | 20 | Zeb, A | Haas, SW | Ozmutlu, S | Hernon, P |
| 6 | Lynch, MF | Lynch, MF | Wolfram, D | *Saracevic, T* | 21 | Mhashi, M | Rada, R | Goodrum, A | Leydesdorff, L |
| 7 | Lunin, LF | Cringean, JK | Furner, J | Zhang, J | 22 | Michailidis, A | Dillon, M | He, SY | Bishop, N |
| 8 | Rada, R | Robertson, AM | Wilson, TD | Wolfram, D | 23 | Mussio, P | *Rousseau, R* | Ozmultu, HC | Rowlands, I |
| 9 | Delia, G | Manson, GA | Foster, A | Furner, J | 24 | Padula, M | *Cool, C* | Furnerhines, J | Tang, R |
| 10 | *Rousseau, R* | *Borgman, CL* | *Saracevic, T* | *Losee, RM* | 25 | Bordogna, G | Meadow, CT | Bookstein, A | *Cool, C* |
| 11 | Lancaster, FW | *Losee, RM* | Jansen, MBJ | Rasmussen, EM | 26 | Carrara, P | Case, DO | Rasmussen, EM | Vakkari, P |
| 12 | Naldi, F | Bookstein, A | Zhang, J | Jarvelin, K | 27 | Bauin, S | Rice, RE | Haas, SW | Abels, EG |
| 13 | Courtial, JP | Meadows, AJ | Jansen, BJ | Thelwall, M | 28 | *Borgman, CL* | Egghe, L | **Wood, FE** | **Wood, FE** |
| 14 | Zimmerman, JL | *Spink, A* | *Cool, C* | *Rousseau, R* | 29 | Laville, F | Lancaster, FW | *Borgman, CL* | Bjorneborn, L |
| 15 | Cooper, M | Lunin, LF | Cole, C | Foster, A | 30 | Vanraan, AFJ | Belkin, NJ | Wilson, T | Vaughan, L |

TABLE 5. Top 30 authors based on betweenness centrality

| Rank | 1988-1992 | 1988-1997 | 1988-2002 | 1988-2007 | Rank | 1988-1992 | 1988-1997 | 1988-2002 | 1988-2007 |
|---|---|---|---|---|---|---|---|---|---|
| 1 | **Willett, P** | *Losee, RM* | *Spink, A* | **Willett, P** | 16 | Glanzel, W | Fox, EA | Iivonen, M | Kretschmer, H |
| 2 | Lunin, LF | *Bookstein, A* | *Losee, RM* | *Spink, A* | 17 | *Bookstein, A* | Abels, EG | *Saracevic, T* | Tang, R |
| 3 | Wood, FE | **Willett, P** | **Borgman, CL** | Chowdhury, GG | 18 | Mussio, P | *Lancaster, FW* | White, MD | **Borgman, CL** |
| 4 | Rada, R | *Spink, A* | Furner, J | Lynch, MF | 19 | Padula, M | Belkin, NJ | Ford, N | Meyer, M |
| 5 | *Bawden, D* | **Rousseau, R** | **Willett, P** | Zhang, J | 20 | Bauin, S | Liebscher, P | Wang, PL | Rowlands, I |
| 6 | Courtial, JP | *Saracevic, T* | *Bookstein, A* | **Rousseau, R** | 21 | Schubert, A | Miquel, JF | Beaulieu, M | Wolfram, D |
| 7 | Naldi, F | Rao, IKR | Zhang, J | *Lancaster, FW* | 22 | Case, DO | Allen, B | Tenopir, C | Vakkari, P |
| 8 | **Rousseau, R** | Beaulieu, M | Ellis, D | Bishop, N | 23 | Meadows, AJ | *Wood, FE* | Oddy, RN | Smith, A |
| 9 | Miquel, JF | **Borgman, CL** | Haas, SW | Hernon, P | 24 | Winterhager, M | Meadow, CT | Bishop, A | *Bawden, D* |
| 10 | *Lancaster, FW* | *Bawden, D* | Korfhage, RR | Ellis, D | 25 | Turner, WA | Tibbo, HR | Mcclure, CR | Fox, EA |
| 11 | Vanraan, AFJ | Meadows, AJ | Myaeng, SH | Thelwall, M | 26 | Dillon, M | Pettigrew, KE | Nahl, D | *Losee, RM* |
| 12 | **Borgman, CL** | Haas, SW | Wolfram, D | *Saracevic, T* | 27 | Woodsworth, A | Cronin, B | Smith, M | Foo, S |
| 13 | Laville, F | Lunin, LF | **Rousseau, R** | Leydesdorff, L | 28 | Braam, RR | Dillon, M | Rao, IKR | Jarvelin, K |
| 14 | Nederhof, AJ | Wood, FE | Meho, LI | Morris, A | 29 | Moed, HF | Kantor, P | Yitzhaki, M | Rasmussen, EM |
| 15 | Egghe, L | Rada, R | Sonnenwald, DH | Kantor, P | 30 | Braun, T | Cool, C | Rice, RE | Furner, J |

TABLE 6. Top 30 authors based on degree centrality

| Rank | 1988-1992 | 1988-1997 | 1988-2002 | 1988-2007 | Rank | 1988-1992 | 1988-1997 | 1988-2002 | 1988-2007 |
|---|---|---|---|---|---|---|---|---|---|
| 1 | **Willett, P** | **Willett, P** | **Rousseau, R** | **Rousseau, R** | 16 | Carrara, P | *Bookstein, A* | *Miquel, JF* | Kostoff, RN |
| 2 | *Rada, R* | **Rousseau, R** | **Willett, P** | **Willett, P** | 17 | Vanraan, AFJ | Beaulieu, M | Choi, KS | Zhang, J |
| 3 | **Rousseau, R** | **Lancaster, FW** | Oppenheim, C | Oppenheim, C | 18 | *Meadows, AJ* | Walker, S | Ellis, D | Glanzel, W |
| 4 | **Lancaster, FW** | *Rada, R* | Chen, HC | Spink, A | 19 | *Bookstein, A* | Vanraan, AFJ | Saracevic, T | Gupta, BM |
| 5 | **Courtial, JP** | **Courtial, JP** | Spink, A | Ford, N | 20 | Lunin, LF | Hancockbeaulieu, M | Gibb, F | Croft, WB |
| 6 | *Wood, FE* | *Meadows, AJ* | **Lancaster, FW** | Leydesdorff, L | 21 | Gagliardi, I | Glanzel, W | *Belkin, NJ* | *Belkin, NJ* |
| 7 | Naldi, F | Padula, M | **Borgman, CL** | **Borgman, CL** | 22 | Merelli, D | Braun, T | Morris, A | Choi, KS |
| 8 | **Bawden, D** | **Borgman, CL** | **Courtial, JP** | **Lancaster, FW** | 23 | Vanhoutte, A | Schubert, A | *Bookstein, A* | Zobel, J |
| 9 | *Miquel, JF* | *Miquel, JF* | *Rada, R* | Jarvelin, K | 24 | Hamers, L | Budd, JM | Robertson, S | Nicholas, D |
| 10 | Mussio, P | *Cronin, B* | Ford, N | Thelwall, M | 25 | Hemeryck, Y | Chen, HC | Croft, WB | Debackere, K |
| 11 | Padula, M | **Bawden, D** | *Cronin, B* | Kantor, P | 26 | Herweyers, G | Woodsworth, A | *Wood, FE* | Miller, D |
| 12 | **Borgman, CL** | *Wood, FE* | *Moed, HF* | *Cronin, B* | 27 | Janssen, M | Haas, SW | Tijssen, RJW | **Bawden, D** |
| 13 | Bauin, S | Saracevic, T | *Meadows, AJ* | *Moed, HF* | 28 | Keters, H | *Belkin, NJ* | Frieder, O | Tenopir, C |
| 14 | Woodsworth, A | *Fox, EA* | Gupta, BM | **Courtial, JP** | 29 | Schubert, A | Allen, B | Wolfram, D | Kelly, D |
| 15 | Bordogna, G | *Moed, HF* | **Bawden, D** | *Fox, EA* | 30 | Lester, J | Hernon, P | *Fox, EA* | Huntington, P |

TABLE 7. Top 30 authors based on PageRank

| Rank | 1988-1992 | 1988-1997 | 1988-2002 | 1988-2007 | Rank | 1988-1992 | 1988-1997 | 1988-2002 | 1988-2007 |
|---|---|---|---|---|---|---|---|---|---|
| 1 | **Willett, P** | **Willett, P** | **Rousseau, R** | *Oppenheim, C* | 16 | Buttlar, L | **Bookstein, A** | Ford, N | Thelwall, M |
| 2 | **Lancaster, FW** | **Lancaster, FW** | **Willett, P** | **Rousseau, R** | 17 | Metz, P | Saracevic, T | *Moed, HF* | **Meadows, AJ** |
| 3 | **Rousseau, R** | **Rousseau, R** | *Oppenheim, C* | **Willett, P** | 18 | Garg, KC | *Croft, WB* | Tenopir, C | *Hernon, P* |
| 4 | *Wood, FE* | *Rada, R* | **Lancaster, FW** | Spink, A | 19 | Yatesmercer, PA | Williams, ME | *Budd, JM* | **Courtial, JP** |
| 5 | *Rada, R* | **Meadows, AJ** | Spink, A | Jarvelin, K | 20 | Schubert, A | Chen, HC | **Bookstein, A** | *Moed, HF* |
| 6 | **Courtial, JP** | **Cronin, B** | Chen, HC | *Leydesdorff, L* | 21 | Bauin, S | *Morris, A* | Harter, SP | *Croft, WB* |
| 7 | **Meadows, AJ** | **Courtial, JP** | **Cronin, B** | **Cronin, B** | 22 | Naldi, F | Voigt, K | *Leydesdorff, L* | Kostoff, RN |
| 8 | **Borgman, CL** | *Budd, JM* | **Meadows, AJ** | **Lancaster, FW** | 23 | *Budd, JM* | Wolfram, D | Wolfram, D | Kling, R |
| 9 | **Bawden, D** | **Borgman, CL** | **Borgman, CL** | Ford, N | 24 | Harris, RM | Frieder, O | Williams, ME | Tenopir, C |
| 10 | **Bookstein, A** | **Bawden, D** | **Courtial, JP** | Zhang, J | 25 | Case, DO | Rice, RE | *Hernon, P* | Mcclure, CR |
| 11 | **Cronin, B** | *Wood, FE* | *Rada, R* | **Borgman, CL** | 26 | Vizinegoetz, D | Delia, G | *Wood, FE* | Choi, KS |
| 12 | Vanraan, AFJ | *Oppenheim, C* | **Bawden, D** | Zobel, J | 27 | Saracevic, T | Vanraan, AFJ | *Croft, WB* | Glanzel, W |
| 13 | Miquel, JF | *Hernon, P* | Gupta, BM | *Morris, A* | 28 | Spangenberg, JFA | *Leydesdorff, L* | Frieder, O | **Bookstein, A** |
| 14 | Pravdic, N | *Moed, HF* | *Morris, A* | Gupta, BM | 29 | Nederhof, AJ | Metz, P | Voigt, K | Connaway, LS |
| 15 | Tague, J | Harter, SP | Ingwersen, P | **Bawden, D** | 30 | Oberg, LR | Dillon, A | Dilevko, J | Fox, EA |

A few authors are consecutively highly ranked through all four time slices between 1988 and 2007. Examples are closeness centrality for Willett, P (1-1-5-2: 1st in 1988-1992, 1st in 1988-1997, 5th in 1988-1997, and 2nd in 1988-2007, the same for rest such format), betweenness centrality for Willett, P (1-3-5-1), betweenness centrality for Borgman, CL (12-9-3-18), betweenness centrality for Rousseau, R (8-5-13-6), degree centrality and PageRank for Willett, P (1-1-2-2; 1-1-2-3), degree centrality and PageRank for Rousseau, R (3-2-1-1; 3-3-1-2), degree centrality and PageRank for Lancaster, FW (4-3-6-8; 2-2-4-8). The twenty years are "golden ages" for these authors: they collaborated frequently (for degree centrality), productively (for PageRank), widely (for closeness centrality), and diversely (for betweenness centrality).

Some authors collaborate more actively in recent years. Spink, A only published one article in 1988-1992 (in this data set), and as a result her centrality for that time slice ranked low, only 224 for closeness centrality, and 797 for degree centrality. Nevertheless, in recent 15 years, she published 53 articles (in this data set), and collaborated with 34 authors, the trends of closeness centrality and degree centrality for her are 224-43-1-1 and 797-105-5-4. Similar situations can also be applied to Ellis, D (closeness centrality: 2054-5-2-3), Saracevic, T (closeness centrality: 170-6-17-12; betweenness centrality: 47-6-17-12), Losee, RM (closeness centrality: 313-11-4-10), Cronin, B (degree centrality: 62-10-11-12), Moed, HF (degree centrality: 175-15-12-13), Fox, EA (degree centrality: 410-14-30-15), Oppenheim, C (PageRank: NA-12-3-1), Leydesdorff, L (PageRank: 58-28-22-6), and Morris, A (PageRank: 44-21-14-13).

Meanwhile, some authors are less collaborative in this field in recent years. Most LIS articles Rada, R published are around 1985-1995; after 1995, his publications are more frequently appeared in computer science journals. Thus, his degree centrality and PageRank is decreasing since then: 2-4-9-1198 for degree centrality and 5-4-11-1850 for PageRank. Most articles Wood, EF published are in the 80s and 90s, and as a result, his centrality rankings are on the decline: 2-3-28-28 for closeness centrality, 3-14-54-168 for betweenness centrality, 6-12-26-69 for degree centrality, and 4-11-26-40 for PageRank. Other examples include Cringean, JK (closeness centrality: 4-7-51-37), Lunin, LF (betweenness centrality: 2-13-137-890), Naldi, F (degree centrality: 8-17-40-532).

A new "force" also rises in this field. Typical example is Thelwall, M: all of his articles are published after 2000, and thus he does not have centrality values for first two time slices and very low values for 1988-2002. Nevertheless, his centrality for 1988-2007 is quite high; all of them are in the top 30: 13th for closeness centrality, 11th for betweenness centrality, 10th for degree centrality, and 16th for PageRank. Other examples include Kelly, D (degree centrality: NA-NA-328-29), Tang, R (closeness centrality: NA-NA-350-24; betweenness centrality: NA-NA-123-17). We can expect that these authors will play a more important role in this field in the coming years.

TABLE 8 lists top 40 authors based on the number of citations to their publications. Corresponding centrality rankings within top 40 are displayed in bold font.

TABLE 8. Top 40 authors based on citation counts

| | Citation* | | Centrality Ranking | | | |
|---|---|---|---|---|---|---|
| Author | Counts | Ranking | Closeness | Betweenness | Degree | PageRank |
| Salton, G | 1464 | 1 | 1199 | 259 | 216 | 229 |
| Buckley, C | 1389 | 2 | 1200 | 260 | 216 | 230 |
| Dumais, ST | 1323 | 3 | 1545 | 172 | 107 | 106 |
| Landauer, TK | 1295 | 4 | 1844 | 382 | 292 | 269 |
| Harshman, R | 1275 | 5 | 1845 | 672 | 554 | 667 |
| Deerwester, S | 1275 | 5 | 1845 | 672 | 554 | 667 |
| Furnas, GW | 1275 | 5 | 1845 | 672 | 554 | 667 |
| Spink, A | 1253 | 8 | **1** | **2** | **4** | **4** |
| Saracevic, T | 1141 | 9 | **6** | **12** | 47 | 84 |
| Glanzel, W | 969 | 10 | 384 | **34** | **18** | **27** |
| Thelwall, M | 884 | 11 | **13** | **11** | **10** | **16** |
| McCain, KW | 835 | 12 | 1432 | 136 | 107 | 103 |
| Ingwersen, P | 791 | 13 | 41 | 74 | 76 | 52 |
| Jansen, BJ | 787 | 14 | **23** | 189 | 62 | 67 |
| Egghe, L | 747 | 15 | 206 | 147 | 107 | 79 |
| Rousseau, R | 705 | 16 | **14** | **6** | **1** | **2** |
| Braun, T | 704 | 17 | 897 | 175 | 47 | 56 |
| Schubert, A | 701 | 18 | 898 | 176 | 47 | 54 |
| Borgman, CL | 685 | 19 | 109 | **18** | **7** | **11** |
| Ellis, D | 654 | 20 | **3** | **10** | **31** | **33** |
| Moed, HF | 639 | 21 | 394 | 63 | **13** | **20** |
| Kantor, P | 635 | 22 | **20** | **15** | **11** | **36** |
| Willett, P | 609 | 23 | **2** | **1** | **2** | **3** |
| White, HD | 608 | 24 | 976 | 115 | 414 | 330 |
| Vanraan, AFJ | 590 | 25 | 728 | 284 | 62 | 85 |
| Cronin, B | 564 | 26 | 353 | **36** | **12** | **7** |
| Harter, SP | 526 | 27 | 1041 | 181 | 136 | 58 |
| Leydesdorff, L | 489 | 28 | **21** | **13** | **6** | **6** |
| Fidel, R | 426 | 29 | 666 | 117 | 47 | 83 |
| Wilson, TD | 414 | 30 | **5** | 44 | 136 | 162 |
| Ford, N | 378 | 31 | **4** | **40** | **5** | **9** |
| Vakkari, P | 361 | 32 | **26** | **22** | 47 | **37** |
| Jarvelin, K | 350 | 33 | **12** | **28** | **9** | **5** |
| Marchionini, G | 346 | 34 | 358 | 41 | **38** | **35** |
| Wolfram, D | 320 | 35 | **8** | **21** | 47 | **32** |
| Oppenheim, C | 295 | 36 | 1969 | 59 | **3** | **1** |
| Large, A | 291 | 37 | 427 | 270 | 41 | 59 |
| Persson, O | 285 | 38 | 402 | 107 | 88 | 98 |
| Losee, RM | 282 | 39 | **10** | **26** | 414 | 346 |
| Kling, R | 274 | 40 | 1262 | 129 | 41 | **23** |

TABLE 8 shows some discrepancies within the rankings of citations and centrality measures. The most obvious one is that the 7 most cited authors have very low centrality rankings. This is due to the fact that they have limited number of papers in our data set (9, 7, 5, 2, 1, 1, and 1 respectively); however, these papers are quite highly cited (Deerwester, S, Dumais, ST, Landauer, TK, Furnas, GW and Harshman, R coauthored a paper been citied 1275 times; Salton, G and Buckley, C coauthored two papers which have been cited 906 and 328 times). As a result, they have very few collaborators (7, 7, 10, 6, 4, 4, and 4 respectively) and most of them are not cut-points (Nooy, Mrvar, & Batagelj, 2005), and accordingly they are in the periphery of the coauthorship network. Some less obvious instances including Ingwersen, P, Jansen, BJ, Marchionini, G and so on, although their centrality rankings correspond to their citation rankings, yet only a portion of their publications are incorporated in our data set, which may affect their ranking results.

Discrepancies also exit within different centrality measures. For example, Glanzel, W has high degree centrality, indicating that he has collaborated with many authors (20 authors), but his closeness centrality is low which ranks only 384 out of 2197. The reason for this is that most of his collaborators locate in Europe, mainly Hungary, Germany and the Netherlands. Thus he is "close" to European authors, whereas distant to authors in other regions, and as a result, his closeness centrality is low. McCain, KW has high citation ranking but low centrality rankings. This is because the author only collaborates with 10 authors and all of her collaborators locate in the USA, thus she does not have high centrality values. The same reasons can also be applied to Ingwersen, Egghe: most of Ingwersen's collaborators locate in Denmark, and most of Egghe's collaborators located in Belgium. By comparison, although the majority of Rousseau's collaborators locate in Belgium, yet he also collaborates with authors from China, Japan, India, England and Canada, thus shortened his virtual distance with authors in the network.

In the interest of gaining a more general perspective of the pattern of collaboration in this coauthorship network, we calculate the Spearman's correlations between centrality measures and citation counts for all authors in the largest component, shown in TABLE 9.

TABLE 9. Spearman's correlations between centrality measures and citation counts

| | Citations | Closeness | Betweenness | Degree | PageRank |
|---|---|---|---|---|---|
| Citations | 1 | 0.2369* | 0.5327* | 0.3964* | 0.4101* |
| Closeness | 0.2369* | 1 | 0.1929* | 0.1983* | 0.1087* |
| Betweenness | 0.5327* | 0.1929* | 1 | 0.6567* | 0.7322* |
| Degree | 0.3964* | 0.1983* | 0.6567* | 1 | 0.9505* |
| PageRank | 0.4101* | 0.1087* | 0.7322* | 0.9505* | 1 |

* Correlation is significant at the 0.01 level.

TABLE 9 shows that four centrality measures have significant correlation with citation counts at the 0.01 level, with betweenness as the highest. The high correlation of citation counts with centrality suggests that centrality measures in certain degree also assess author's scientific productivity and quality. They can be indicators, or at least supplementary indicators for impact evaluation, providing alternative perspectives for current methods.

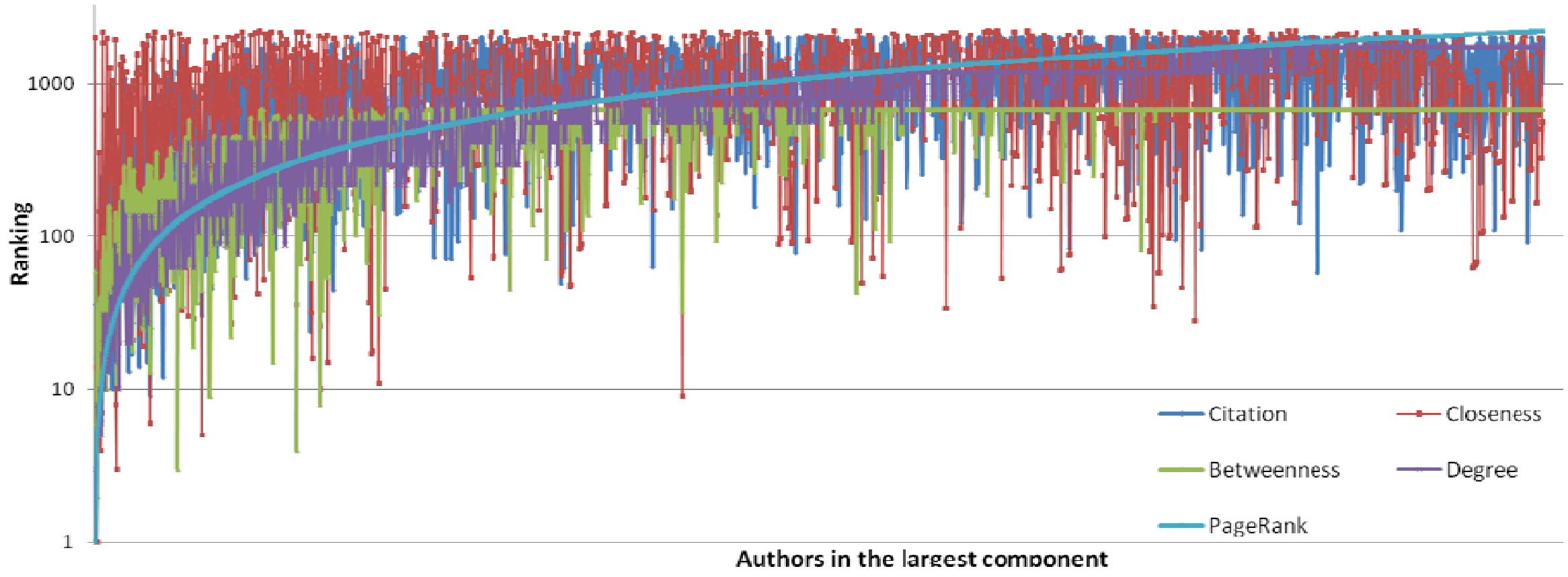

FIG. 3. Distribution of rankings of citation counts and centrality measures.

FIG. 3 shows the distribution of rankings of citation counts and centrality measures. X axis stands for all the authors in the largest component ranked by PageRank, Y axis stands for rankings, from 1st to 2197th. The reason we choose PageRank as the benchmark is that PageRank have relatively highest correlations with rest centrality measures, and thus improving the readability and consistency of the graph. From FIG. 3 we can discover that the overall distribution of the ranking of citation counts matches that of centrality measures, which is in accordance with the results shown in Spearman's correlations. Rankings of PageRank, degree centrality and betweenness centrality correlate with each other more precisely; while rankings of citation counts and closeness centrality have some inconsistent values.

## 5. Discussion and Conclusion

The evolving coauthorship network is effective in revealing the dynamic collaboration patterns of authors. The different positions authors belong to at each time slice reflect the collaboration trend of authors. We find that some authors are consecutively highly ranked in all time periods, indicating that these they are on the "plateau" of their academic career; comparatively, some authors are on the rise in this field while some are faded out.

We also verify the correlation between citation and centrality. We find that all the four centrality measures are significantly correlated with citation counts, whereas some inconsistencies occur. The discrepancy can be interpreted from two perspectives. First, citations and centralities measure different contents. Although the motivation for citation varies, citation counts measure the quality and impact of articles (Garfield & Sher, 1963; Frost, 1979; Lawani & Bayer, 1983; Baird & Oppenheim, 1994). While centrality measures both article impact and author's field impact. Degree centrality measures author's collaboration scope, closeness centrality measures author's position and virtual distance with others in the field, and betweenness centrality measures author's importance to other authors' virtual communication. Hence, centrality has its value in impact evaluation, since it integrates both article impact and author's field impact. Their relationship can be illustrated in FIG. 4.

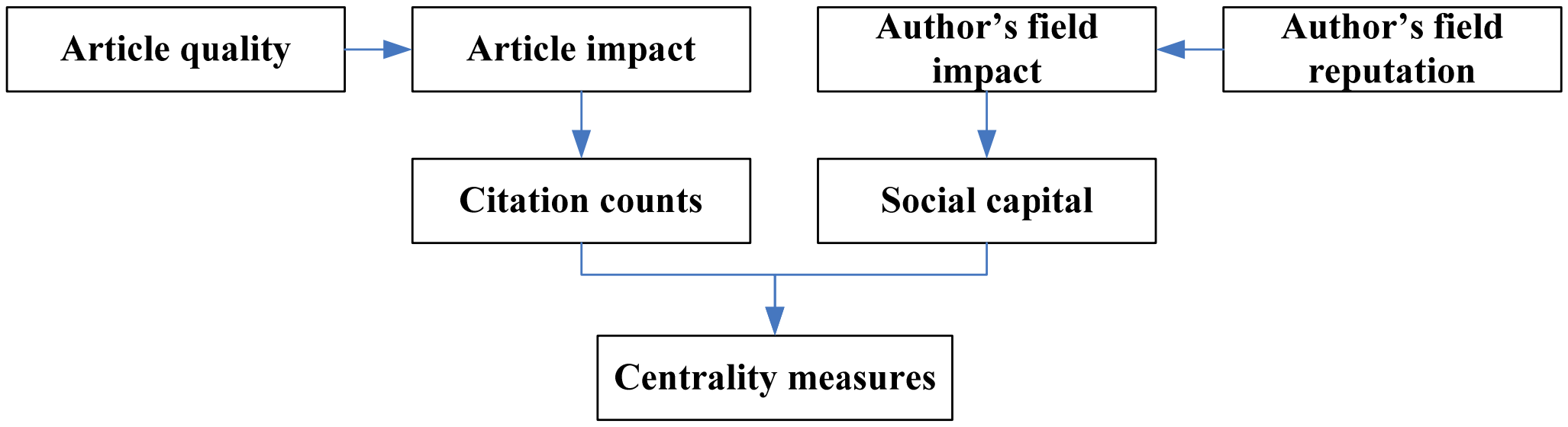

FIG. 4. Relation between citation and centrality

The quality of an article is subjective, yet we can measure it indirectly through article impact which can be quantified by citation counts. Similar to article quality, author's field reputation is also difficult to assess, but we can measure it through social capital (Burt, 1980; Burt, 2002; Cronin & Shaw, 2002). Accordingly, centrality measures integrate both article impact – citation counts and author's field impact– social capital, as displayed in FIG. 4.

Another factor contributed to these discrepancies is the limitations inherent to current algorithm of centrality measures. Authors from papers coauthored by multiple authors have high degree centrality. This may be magnified when coauthored with many authors. For instance, if a paper is coauthored by 10 authors, each of these authors would have a degree centrality of 9. This is equivalent to 45 papers if they were coauthored by just two authors. It is obvious that they have quite different academic impacts. Closeness centrality is a measure of network property rather than a direct measure of academic impact. Any author coauthoring an article with authors having high closeness centrality would also result in a high closeness centrality; however, this author may have little academic impact. Authors involved in interdisciplinary research would have a high betweenness centrality even through their role in this specific discipline may not be that significant. Centrality measures will be much more useful and valuable if these drawbacks have been eliminated.

In fact, some scholars have already embarked on this. Newman (2005) proposed a new betweenness measure that includes contributions from essentially all paths between nodes, not just the shortest, and meanwhile giving more weight to short paths. Brandes (2008) introduced variants of betweenness measures, as endpoint betweenness, proxies betweenness, and bounded distance betweenness. Liu et al. (2005) defined AuthorRank, a modification of PageRank which considers link weight. Other work aiming at improving PageRank in the context of author ranking includes Sidiropoulos and Manolopoulos (2005), and Fiala, Rousselot and Ježek (2008).

In future studies, it will be necessary to improve the algorithm of centrality measure, and utilize their strength in improving the current impact evaluation. Potentially, it is possible and necessary to apply centrality measures to other social networks (e.g. co-citation networks), and add semantics to them (e.g. Mesur Ontology and SWRC Ontology), and thus construct a systematic model for evaluation indicators.

## References

Albert, R., & Barabási, A. (2002). Statistical mechanics of complex networks. *Review of Modern Physics, 74*(1), 47-97.

Baird, L. M., & Oppenheim, C. (1994). Do citations matter? *Journal of Information Science, 20*(1), 2-15.

Barabási, A. L. (2003). *Linked - how everything is connected to everything else and what it means for business, science, and everyday life*. New York: Plume.

Barabási, A. L., Jeong, H., Neda, Z., Ravasz, E., Schubert, A., & Vicsek, T. (2002). Evolution of the social network of scientific collaborations. *Physica A, 311*(3-4), 590-614.

Boje, D. M., & Whetten, D. A. (1981). Effects of organizational strategies and contextual constraints on centrality and attributions of influence in inter-organizational networks. *Administrative Science Quarterly, 26*(3), 378-395.

Bollen, J., Rodriguez, M. A., & Van De Sompel, H. (2006). Journal status. Scientometrics, 69(3), 669-687.

Bonacich, P. (1987). Power and centrality: a family of measures. *The American Journal of Sociology, 92*(5), 1170-1182.

Brandes, U. (2008). On variants of shortest-path betweenness centrality and their generic computation. *Social Networks, 30*, 136-145.

Burt, R. S. (1980). Autonomy in a social topology. *American Journal of Sociology, 85*, 892-925.

Burt, R. S. (2002). The Social Capital of Structural Holes. In M. F. Guillén, R. Collins, P. England, & M. Russell (Ed.), *New Directions in Economic Sociology* (pp. 203-247). Thousand Oaks, CA: Sage Foundation.

Chen, C. M. (2006). CiteSpace II: Detecting and visualizing emerging trends and transient patterns in scientific literature. *Journal of the American Society for Information Science and Technology, 57*(3), 359-377.

Cronin, B., & Shaw, D. (2002). Banking (on) different forms of symbolic capital. *Journal of the American Society for Information Science and Technology, 53*(14), 1267-1270.

Crosbieburnett, M. (1984). The centrality of the step relationship - a challenge to family theory and practice. *Family Relations, 33*(3), 459-463.

Dellavalle, R. P, Schilling, L. M., Rodriguez, M. A., Van de Sompel, H, & Bollen, J. (2007). Refining dermatology journal impact factors using PageRank. *Journal of the American Academy of Dermatology, 57*(1), 116-119.

Estrada, E., & Rodriguez-Velazquez, J. A. (2005). Subgraph centrality in complex networks. *Physical Review E, 71*(5), 056103.

Everett, M. G., & Borgatti, S. P. (1999). The centrality of groups and classes. *Journal of Mathematical Sociology, 23*(3), 181-201.

Farmer, T. W., & Rodkin, P. C. (1996). Antisocial and prosocial correlates of classroom social positions: The social network centrality perspective. *Social Development, 5*(2), 174-188.

Fiala, D., Rousselot, F., & ježek, K. (2008). PageRank for bibliographic networks. *Scientometrics, 76*(1), 135-158.

Freeman, L.C. (1977). A set of measures of centrality based on betweenness. *Sociometry, 40*(1), 35-41.

Freeman, L.C. (1979). Centrality in social networks. Conceptual clarification. *Social Networks, 1*, 215-239.

Friedkin, N. E. (1991). Theoretical foundations for centrality measures. *The American Journal of Sociology, 96*(6), 1478-1504.

Frost, C. O. (1979). The use of citations in literary research: a preliminary classification of citation functions. *Library Quarterly, 49*(4), 399-414.

Garfield, E. (1983). *Citation indexing - its theory and application in science, technology and humanities*. Philadelphia: ISI Press.

Hackman, J. D. (1985). Power and centrality in the allocation of resources in colleges and universities. *Administrative Science Quarterly, 30*(1), 61-77.

Ibarra, H. (1993). Network centrality, power, and innovation involvement - determinants of technical and administrative roles. *Academy of Management Journal, 36*(3), 471-501.

Ibarra, H., & Andrews, S. B. (1993). Power, social-influence, and sense making - effects of network centrality and proximity on employee perceptions. *Administrative Science Quarterly, 38*(2), 277-303.

Kretschmer, H. (2004). Author productivity and geodesic distance in bibliographic co-authorship networks and visibility on the Web. *Scientometrics, 60*(3), 409-420.

Lawani, S. M., & Bayer, A. E (1983). Validity of citation criteria for assessing the influence of scientific publications - new evidence with peer assessment. *Journal of the American Society for Information Science, 34*(1), 59-66.

Leydesdorff, L. (2007). Betweenness centrality as an indicator of the interdisciplinarity of scientific journals. *Journal of the American Society for Information Science and Technology, 58*(9), 1303-1319.

Liu, L. G., Xuan, Z. G., Dang, Z. Y., Guo, Q., & Wang, Z. T. (2007). Weighted network properties of Chinese nature science basic research. *Physica A-Statistical Mechanics and Its Applications, 377*(1), 302-314.

Liu, X., Bollen, J. Nelson, M. L., & Sompel, H. V. (2005). Co-authorship networks in the digital library research community. *Information Processing and Management, 41*, 1462-1480.

Mutschke, P. (2003). Mining networks and central entities in digital libraries. A graph theoretic approach applied to co-author networks. *Advances In Intelligent Data Analysis V, 2810*, 155-166.

Nascimento, M. A., Sander, J., & Pound, J. (2003). Analysis of SIGMOD's coauthorship graph. *SIGMOD Record, 32*(3), 8-10.

Newman, M. E. J. (2001a). Scientific collaboration networks: I. Network construction and fundamental results. *Physical Review E, 64*, 016131.

Newman, M. E. J. (2001b). The structure of scientific collaboration networks. *Proceedings of the National Academy of Science of the United States of America, 98*(2), 404-409.

Newman, M. E. J. (2003). The structure and function of complex networks. *SIAM Review, 45*(2), 167-256.

Newman, M. E. J. (2005). A measure of betweenness centrality based on random walks. *Social Networks, 27*, 39-54.

Nisonger, T.E., & Davis, C.H. (2005). The perception of library and information science journals by LIS education deans and ARL library directors: A replication of the Kohl–Davis study. *College & Research Libraries, 66*, 341–77.

Nooy, W., Mrvar, A., & Batagelj, V. (2005). *Exploratory social network analysis with pajek*. Cambridge, UK: Cambridge University Press.

NWB Team. (2006). Network Workbench Tool. Indiana University, Northeastern University, and University of Michigan. Retrieved on November, 21, 2008 from http://nwb.slis.indiana.edu

Page, L., & Brin, S. (1998). The anatomy of a large-scale hypertextual Web search engine. *Computer Networks and ISDN Systems, 30*, 107-117.

Paullay, I. M., Alliger, G. M., & Stoneromero, E. F. (1994). Construct-validation of 2 instruments designed to measure job involvement and work centrality. *Journal of Applied Psychology Volume, 79*(2), 224-228.

Price, J. D. S. (1965). Networks of scientific papers. *Science, 149*, 510-515.

Rodriguez, M. A., & Pepe, A. (2008). On the relationship between the structural and socioacademic communities of a coauthorship network. *Journal of Informetrics, 2*(3), 195-201.

Sidiropoulos, A., & Manolopoulos, Y. (2006), A Generalized comparison of graph-based ranking algorithms for publications and authors. *Journal of Systems and Software, 79*(12), 1679-1700.

Stryker, S., & Serpe, R. T. (1994). Identity salience and psychological centrality - equivalent, overlapping, or complementary concepts. *Social Psychology Quarterly, 57*(1), 16-35.

Verplanken, B., & Holland, R. W. (2002). Motivated decision making: Effects of activation and self-centrality of values on choices and behavior. *Journal of Personality and Social Psychology Volume, 82*(3), 434-447.

Vidgen, R., Henneberg, S., & Naude, P. (2007). What sort of community is the European Conference on Information Systems? A social network analysis 1993-2005. *European Journal of Information Systems, 16*(1), 5-19.

Wasserman, S., & Faust, K. (1994). *Social network analysis*. Cambridge, UK: Cambridge University Press.

Yin, L., Kretschmer, H., Hanneman, R. A., & Liu, Z. (2006). Connection and stratification in research collaboration: An analysis of the COLLNET network. *Information Processing and Management, 42*, 1599-1613.